\documentclass[onecolumn,draftclsnofoot]{IEEEtran}

\usepackage{times,amsmath,epsfig,latexsym,amssymb,psfrag,booktabs,extpfeil}

\begin{document}

\title{Power Allocation in Multi-hop OFDM Transmission Systems with Amplify-and-Forward Relaying: A Unified Approach }

\author{Amin~Azari,~\IEEEmembership{}~Jalil~S.~Harsini,~\IEEEmembership{} and~Farshad~Lahouti,~\IEEEmembership{Member,~IEEE}}

\maketitle

\begin{abstract}
In this paper a unified approach for power allocation (PA) problem
in multi-hop orthogonal frequency division multiplexing (OFDM)
amplify-and-forward (AF) relaying systems has been developed. In the
proposed approach, we consider short and long-term individual and
total power constraints at the source and relays, and devise
decentralized low complexity PA algorithms when wireless links are
subject to channel path-loss and small-scale Rayleigh fading. In
particular, aiming at improving the instantaneous rate of multi-hop
transmission systems with AF relaying, we develop (i) a near-optimal
iterative PA algorithm based on the exact analysis of the received
SNR at the destination; (ii) a low complexity-suboptimal iterative
PA algorithm based on an approximate expression at high-SNR regime;
and (iii) a low complexity-non iterative PA scheme with limited
performance loss. Since the PA problem in multi-hop systems is too
complex to solve with known optimization solvers, in the proposed
formulations, we adopted a two-stage approach, including a power
distribution phase among distinct subcarriers, and a power
allocation phase among different relays. The individual PA phases
are then appropriately linked through an iterative method which
tries to compensate the performance loss caused by the distinct
two-stage approach. Simulation results show the superior performance
of the proposed power allocation algorithms.
\end{abstract}

\section{Introduction}
\newcommand{\ud}{\mathrm{d}}
\IEEEPARstart{M}{ulti-hop} relaying and orthogonal frequency
division multiplexing (OFDM) are promising techniques for high-speed
data communication among wireless devices that may not be within
direct transmission range of each other. Relaying protocols are
broadly categorized as amplify-and-forward (AF) relaying, in which
each relay forwards a scaled version of the received noisy copy of
the source signal, and decode-and-forward (DF) relaying, in which
each relay forwards a regenerated version of the received noisy copy
of the source signal. AF relays may also be categorized as
blind/fixed gain, channel assisted, and channel noise assisted,
based on how source-relay channel state information (CSI) and noise
statistics affect the relay gains \cite{ca:cna}. The capacity
analysis and transmission protocol design over relay channels have
attracted lots of research activities in the past decade
\cite{kram}- \nocite{nosrat}  \cite{lan}. As
relays have shown their merit for data transfer purposes, multi-hop
communications have been also included in advanced wireless
standards such as \uppercase{IEEE} 802.11n \cite{ieee-new}, WiMAX, and LTE-Advanced \cite{ieee-ano}- \nocite{mul:cho} \nocite{ltea}\cite{lteee}. To achieve power efficiency in multi-hop transmission systems, it is necessary to devise efficient power allocation (PA) strategies
for the source and intermediate relays, when a multi-hop data
transmission path is setup. For the simplest form of dual-hop
relaying systems, the PA problem has been investigated in
\cite{paaf}-\nocite{jmimo}\nocite{path}\nocite{sspa}
\nocite{ajsrpa}\cite{ont:iha}. Specifically in \cite{ont:iha}, for a
two-hop OFDM communication system with a given power budget, the
authors presented the optimal power allocation at the relay (source) node for a
given source (relay) power allocation scheme, that maximizes the instantaneous rate
of the system. Then, using simulations they showed that by iterative
power allocation between source and relay, a higher gain is achieved. 
A jointly optimal subcarrier pairing and power allocation scheme which maximizes the  throughput of OFDM amplify-and-forward relaying systems subject to a statistical delay constraint is investigated in \cite{new3}.
Power allocation for dual-hop OFDM relaying systems has been also
considered in \cite{pacpa}-\nocite{agfw}\nocite{ajoipa}
\nocite{ssrpa}\nocite{wen}\nocite{opsoppa}\cite{new1}.     
The power allocation problem for
relaying system models with more than two hops over narrowband
fading channels has been considered in \cite{opt:moh}, \cite{poaf}.
Especially in \cite{poaf}, aiming at maximizing the instantaneous
rate, the power allocation solution for AF relaying protocol over narrowband
Rayleigh fading channels has been provided. In \cite{new2} a path searching algorithm has been presented to find the best links among relays and then two subcarrier allocation algorithms are presented which aim at resource utilization improvement. 
Optimal and suboptimal
power allocation schemes for multi-hop OFDM systems with DF relaying
protocol has been developed in \cite{osopa}. Adaptive power
allocation algorithms for maximizing system capacity (when full CSI
is available), and minimizing system outage probability (with
limited CSI) have been proposed in \cite{apafor} for multi-hop DF
transmission systems under a total power constraint. Aiming at
maximizing the end-to-end average transmission rate under a
long-term total power constraint, the authors in \cite{e2epa2}
developed a resource allocation scheme for multi-hop OFDM DF
relaying system in which the power allocated to each subcarrier and
the transmission time per hop have been specified. 
In \cite{apafmh},
the authors proposed the solution for power allocation problem in multi-hop OFDM
relaying system (AF and DF) under total short-term power constraint,
where the PA and capacity analysis is developed based on a high SNR
approximation in the AF relaying protocol. The approximation used in
\cite{apafmh} has been originally proposed in \cite{out:moh} and
performs well for small number of hops with high received SNR. In
low-to-medium SNR regime or for multi-hop systems with more than
three hops, this approximation loses its functionality in design of
PA schemes. The multi-hop OFDM transmission system has been also
considered in \cite{hajagha}, where joint power allocation and
subcarrier pairing solutions for AF and DF relaying protocols have
been devised under short-term total power constraint. The analysis
for AF relaying protocol in \cite{hajagha} is also based on the high
SNR approximation as discussed above.

\subsection{Paper's contributions}
Although several research works have been reported on the power allocation problem
in multi-hop OFDM systems, to the best of the author's knowledge, no
attempt has been made to develop a unifying approach addressing
different aspect of these systems. In this paper, we focus on
multi-hop OFDM systems with AF relaying and different system power
constraints. We developed an unified framework for efficient power allocation
which includes both iterative and non-iterative solutions. In
particular, aiming at maximizing the instantaneous system rate under
individual and total short and long-term power constraints, the
following power allocation schemes have been devised: (i) A near-optimal iterative
PA algorithm which is developed based on the analysis of an exact
expression for the received SNR at the destination; (ii) A low
complexity-suboptimal iterative PA algorithm in which we use an
high-SNR approximation of the system rate for design purposes;  and
(iii) A low complexity-non iterative PA scheme based on a high-SNR
rate analysis at the destination. The rest of this paper is
organized as follows. In Section II we introduce the system model
and the power allocation problem formulation. An iterative PA solution based on
the analysis of the exact destination SNR is presented in Section
III. In Section IV we provided sub-optimal PA schemes based on the
analysis of an approximated high SNR expression at the destination.
Simulation results are provided in Section IV, and concluding
remarks are presented in Section V.

\providecommand{\abs}[1]{\lvert#1\rvert}

\section{System Model and Problem Formulation }\label{broad}
Fig. \ref{fig1} shows the multi-hop transmission system
model under consideration, where OFDM is utilized for broadband communication among
the consecutive nodes. We assume that the system uses a routing algorithm, and therefore the path between the source and destination nodes is already established. Here the source node, $T_0$, sends data bits to the destination node $T_N$ via $N$-1 intermediate relay nodes,
$T_1,T_2, \ldots,T_{N-1}$, over orthogonal time slots and orthogonal
subcarriers. The fading gain of the narrowband subchannel $i$
(corresponding to the $i$th subcarrier) between nodes $T_{k-1}$ and
$T_k$, denoted by $a_{k,i}$, is modeled as a zero mean circularly
symmetric complex Gaussian random variable with variance
$\sigma_{k,i}^2$. In an AF multi-hop relaying system, each relay
first amplifies the signal received from its immediate preceding
node, and then forwards it to the next node in the subsequent time
slot. The amplification gain in $i$th subcarrier of node $T_k$ is
adapted based on the instantaneous fading amplitude of the channel
between nodes $T_{k-1}$ and $T_k$, i.e. $\abs{ a_{k,i}}$. To ensure
an output relay transmit power $P_{k,i}$ on $i$th subcarrier, the
amplification gain is adjusted as \cite{ont:iha}:
\begin{figure}[!t]
  \includegraphics[width=3.5in]{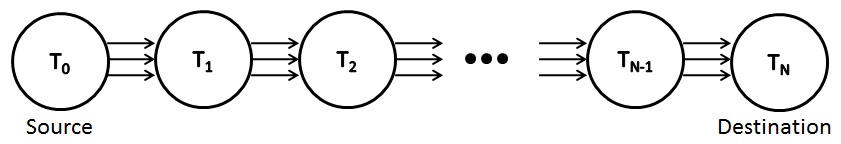}
  \caption{Multi-hop OFDM relaying system with $N$-1 intermediate relays.}\label{fig1}
\end{figure}
\begin{equation}
\begin{array}{l}
A_{{k,i}}^2=\frac
{{P}_{{k,i}}}{{P}_{{k-1,i}}{\abs{a_{{k,i}}}^2}+N_{{0}_{k,i}}}
\; \nonumber
\hspace{6mm}\text{for}\hspace{4mm}   k=1, \ldots,N-1; i=1, \ldots,N_{F}
\end{array}
\end{equation}
where $P_{{0},i}$ and $P_{k,i}$ denote the transmission powers at
source and $k$th relay for $i$th subcarrier, respectively. In this
model, the number of subcarriers, i.e., the number of points for fast
Fourier transform (FFT), and the noise  power at $k$th relay
within $i$th subcarrier are denoted by $N_{F}$ and $N_{{0}_{k,i}}$ respectively.

For the considered multi-hop OFDM system model, the instantaneous received SNR
over the $i$th subcarrier at the destination node is given by
\begin{equation}\label{sntt36}
\begin{array}{l}
\gamma_{T}^i=\left(\prod_{k=1}^N
(1+\frac{1}{\tilde{\gamma}_{k,i}})-1\right)^{-1},
\end{array}
\end{equation}
where $\tilde{\gamma}_{k,i} =
{\frac{P_{k-1,i}}{{N_{{0}}}_{k,i}}}\abs{{a_{k,i}}}^2$ denotes the
instantaneous received SNR over the $i$th subcarrier of $k$th hop
with the average $\tilde{\Gamma}_{k,i} =
{\frac{P_{k-1,i}}{{N_{{0}}}_{k,i}}}{{\sigma_{k,i}^2}}$.
Using \eqref{sntt36}, the instantaneous rate of the
end-to-end multi-hop system is given by
\begin{equation}\label{Ccap}
\begin{array}{l}
C=\frac{1}{N}\sum_{i=1}^{N_{F}}{\log(1+\gamma_{T}^i)}.
\end{array}
\end{equation}
Considering $\alpha_{k,i}=\frac{P_{k,i}}{P_T}$ and ${\gamma}_{k,i} =
\frac{P_{T}{\abs{a_{k,i}}}^2}{{N_{{0}}}_{k,i}}$, $C$ may be rewritten as follows
\begin{equation}\label{Ccap1}
\begin{array}{l}
C=\frac{1}{N}\sum_{i=1}^{N_{F}}{\log\left(1+\frac{1}{\prod_{k=1}^N
(1+\frac{1}{{\alpha_{k,i}\gamma}_{k,i}})-1}\right)}.
\end{array}
\end{equation}
In a Rayleigh fading environment, ${\gamma}_{k,i}$ follows an exponential
distribution with the average ${\Gamma}_{k,i} =
\frac{P_{T}{\sigma_{k,i}}^2}{{N_{{0}}}_{k,i}}$.

Given the power constraint, here the goal is to find the
transmit powers of the subcarriers at the source and the relay nodes
such that the instantaneous rate in \eqref{Ccap1} is maximized. In
this paper, we consider power allocation optimization problem under short-term
power constraints (STPC), long-term individual power constraints
(LTIPC), and long-term total power constraints (LTTPC). The general form of power allocation optimization (PA) problem in this work is as follows.
\begin{equation}\label{op2op0}
\begin{array}{l}
\hspace{5mm}\max_{\alpha_{k,i}} \quad C \qquad \mathrm{s.t.} \\
\quad\sum_{i=1}^{N_{F}}\sum_{k=0}^{N-1}\alpha_{k,i}=1  \hspace{18mm}  \text{STPC}\\
\\
  \quad\mathbb{E}(\alpha_{k,i})=\frac{1}{N \times N_{F}}  \hspace{24mm}  \text{LTIPC}\\
\\
 \quad\sum_{i=1}^{N_{F}}\sum_{k=0}^{N-1}\mathbb{E}(\alpha_{k,i})=1  \hspace{13mm}  \text{LTTPC}\\
\end{array}
\end{equation}
To explicitly identify the way power is distributed among different subcarriers and nodes, we denote by $P_{i}$ and $P_{k,i}$, respectively, the allocated power to $i$th subcarrier
(for all nodes) and the allocated power to $i$th subcarrier in $k$th
node (hop). We also define two new nonnegative PA coefficients
$\mu_{i}$ and $\beta_{k,i}$ as follows
\begin{align}
P_{i}=&\mu_{i}\times P_{T}\\ 
\text{and} \quad P_{k,i}=&\beta_{k,i}\times P_i=\beta_{k,i}\times\mu_{i}\times P_{T},
\end{align}
where $\alpha_{k,i}=\beta_{k,i}\times\mu_i$. The optimization problem (3) may now be rewritten as follows with $\mu_{i}$ and $\beta_{k,i}$ as optimization variables,
\begin{equation}\label{op2op1}
\begin{array}{l}
\hspace{5mm}\max_{\beta_{k,i},\mu_i} \quad C \qquad \mathrm{s.t.} \\
\mathrm{C1.} \quad\\
 \hspace{6mm}\sum_{i=1}^{N_{F}}\sum_{k=0}^{N-1}\mu_i\beta_{k,i}=1  \hspace{25mm} \text{STPC}\\
 \hspace{6mm} \mathbb{E}(\mu_{i})=\frac{1}{N_{F}}, \quad \mathbb{E}(\beta_{k,i})=\frac{1}{N}\hspace{17mm} \text{LTIPC}\\
 \hspace{6mm} \sum_{i=1}^{N_{F}}\mathbb{E}(\mu_{i})=1 \quad \sum_{k=0}^{N-1}\mathbb{E}(\beta_{k,i})=1  \hspace{2mm}\text{LTTPC}\\
\mathrm{C2.} \quad \mu_i \geq 0 \quad  i= 1,\ldots, N_{F}\\
\mathrm{C3.} \quad \beta_{k,i} \geq 0 \quad  k= 0,\ldots, N-1 ;  i= 1,\ldots, N_{F}.
\end{array}
\end{equation}

We note that the allocation of power to subcarriers (by finding
$\mu_i$) and to subcarrier per node (by finding $\beta_{k,i}$) in
problem (6) is equivalent to finding $\alpha_{k,i}$ in (3).
Unfortunately, the power allocation problems (3) and (6) are too complicated and
cannot be solved with known optimization solvers. Hence, in the
subsequent sections to find efficient solutions to \eqref{op2op1}
under different power constraints, we develop iterative and
non-iterative algorithms based on the exact and approximate SNR expressions. \\

\section{Iterative Power Allocation}
In this Section, we consider the power allocation problem (6) as alternate maximization over two
simpler power allocation optimization sub-problems. In the first sub-problem, the optimized $\{\beta_{k,i}\}$ is obtained assuming that $\{\mu_{i}\}$ is available. In the second sub-problem, $\{\mu_{i}\}$ is determined for a given set of $\{\beta_{k,i}\}$. Next, we provide an iterative PA
algorithm in which the two subproblems are alternately considered with the output of one as the input of the other. Numerical evaluation in Section V verifies the effectiveness of the proposed approach. 

\subsection{Sub-problem 1: Power allocation among relays}
Given that the power allocated to each of the subcarriers $\{\mu_i\}$ is already known, the power allocation problem among relays in one subcarrier is equivalent to the power allocation problem in a
multi-hop system with narrowband fading model. Here, we provide
optimal solutions for the PA problems in multi-hop narrowband
communication systems with  individual long-term and total power
constraints. We note that in an AF multi-hop transmission system, maximization of the
instantaneous rate of the system is equivalent to maximizing the instantaneous received SNR \cite{poaf}. The instantaneous received SNR at the destination of multi-hop
system given in \eqref{sntt36} may be expressed as follows:

\begin{equation*}
\begin{array}{l}
\ln\left(1+{\gamma_{T}^i}^{-1}\right)=\ln\prod_{k=1}^N(
1+\frac{1}{\tilde{\gamma}_{k,i}})=\sum_{k=1}^N
\ln(1+\frac{1}{{\tilde{\gamma}_{k,i}}}).
\end{array}
\end{equation*}
Then, $\gamma_T^i$ is found as
\begin{equation}\label{Gamma_t}
\begin{array}{l}
\gamma_{T}^i=\left[\exp \left(\sum_{k=1}^N\ln(1+\frac{1}{\tilde{\gamma}_{k,i}})\right)-1\right]^{-1}.
\end{array}
\end{equation}
Since, maximizing $\gamma_{T}^i$ is equivalent to minimizing the
argument of exponential function on the right hand side of 
\eqref{Gamma_t}, we can conclude that
\begin{equation}\label{newmax}
\begin{array}{l}
\text{arg}\max_{\{\beta_{k,i}\}}\gamma_{T}^i =  \text{arg}\min_{\{\beta_{k,i}\}}{\sum_{k=1}^N \ln(1+\frac{1}{\tilde{\gamma}_{k,i}})}.
\end{array}
\end{equation}

Using \eqref{newmax}, the optimization problem for the $i$th
subcarrier under LTIPC is written
as follows:

\begin{equation}\label{optschaver1}
\begin{array}{l}
\hspace{5mm} \min_{\{\beta_{k,i}\}} \quad  \sum_{k=0}^{N-1} \ln(1+\frac{1}{\beta_{k,i}\mu_i\gamma_{k+1,i}}) \\
\\
\mathrm{s.t.} \quad \mathbb{E}(\beta_{k,i})=\frac{1}{N},\quad k=0, 1, \ldots,
N-1,
\end{array}
\end{equation}
where  $\{\mu_i\}$ is given. In appendix
\ref{ap1conv}, it is shown that the objective function in
\eqref{optschaver1} is convex. Since the constraints are linear the
main problem is convex, and therefore we use the Lagrange
method to obtain the optimal solution. The Lagrangian function is given by
\begin{equation}
\begin{array}{l}
\mathcal{L}= \sum_{k=0}^{N-1} \ln(1+\frac{1}{\beta_{k,i} \mu_i\gamma_{k+1,i}})
+\sum_{k=0}^{N-1} \lambda_k\left(\mathbb{E}(\beta_{k})-\frac{1}{N}\right)\\
\end{array}
\end{equation}
where $\{ \lambda _{k,i} \}_{k = 0}^{N - 1}$ are Lagrange
multipliers. By taking the derivative of $\mathcal{L}$ with respect
to $\beta_{k,i},k = 0,1,...,N - 1$, one obtains
\begin{equation}\label{betaop}
\begin{array}{l}
\frac{\partial\mathcal{ L}}{\partial \beta_{k,i}}=0 \quad \rightarrow \quad
\lambda_{k,i}-\frac {1}{\beta_k(\mu_i\gamma_{k+1,i}\beta_{k,i})} =0 \\
\\
\hspace{12mm}\rightarrow\beta_{k,i}=\frac{-1+\sqrt{1+4\frac{\mu_i\gamma_{k+1,i}}{\lambda_{k,i}}}}{2\mu_i\gamma_{k+1,i}}.
\end{array}
\end{equation}
Substituting \eqref{betaop} in the $k$th individual power constraint
in (10) leads to
\begin{equation}\label{16}
\begin{array}{l}
\mathbb{E}(\beta_{k,i})=\frac{1}{N} \quad \rightarrow \quad
\mathbb{E}\left(\frac{-1+\sqrt{1+4\frac{\mu_i\gamma_{k+1,i}}{\lambda_{k,i}}}}{2\mu_i\gamma_{k+1,i}}\right)=\frac{1}{N}\\
\\
\hspace{6mm}\rightarrow\int_{{0}}^{\infty}
\frac{\sqrt{1+4\frac{\mu_i\gamma_{k+1,i}}{\lambda_{k,i}}}-1}{2\mu_i\gamma_{k+1,i}}\frac{
e^{\frac{-\mu_i\gamma_{k+1,i}}{\mu_i\Gamma_{k+1,i}}}}{\mu_i \Gamma_{k+1,i}}{d}_{\gamma_{k+1,i}}=\frac{1}{N}.
\end{array}
\end{equation}

To find the Lagrange multiplier $\lambda_{k,i}$,  the integral in
\eqref{16} can be numerically evaluated and a
bisection root-finding method \cite{XX} can be utilized. From
\eqref{betaop} and \eqref{16}, it is evident that the power
coefficient at the $k$th node, $\beta_{k,i}, k = 0, \ldots, N - 1$,
is only dependent on the fading gain of its immediate forward
channel. As a result, the proposed power allocation scheme can be implemented in a
decentralized manner. Such a PA scheme is potentially applicable in
ad-hoc wireless networks over narrowband channels. 

Using similar steps for the LTTPC case, the PA coefficient is obtained by
\begin{equation}
\begin{array}{l}
\beta_{k,i}=\frac{-1+\sqrt{1+4\frac{\mu_i\gamma_{k+1,i}}{\lambda_{i}}}}{2\mu_i\gamma_{k+1,i}},
\end{array}
\end{equation}
where the constant $\lambda$ is calculated using the corresponding power constraint
follows
\[ \sum_{k=0}^{N-1}\mathbb{E}(\beta_{k,i})=1. \]
Moreover, for the STPC case, the procedure is similar to the LTIPC
scenario.

\subsection{Sub-problem 2: Power allocation among subcarriers}
The instantaneous rate, $C$, can be written as a function of $\mu_i$s and $\beta_{k,i}s$, as follows
\begin{equation}\label{casl}
\begin{array}{l}
C=\frac{1}{N}\sum_{i=1}^{N_{F}}{\log\left(1+\frac{1}{\prod_{k=0}^{N-1}
(1+\frac{1}{{\beta_{k,i}\mu_i\gamma}_{k+1,i}})-1}\right)}.
\end{array}
\end{equation}
Given $\beta_{k,i}$s, $C$ is to be maximized by
finding the optimal $\mu_i$s. We start by
expanding the expression for $\gamma_T^i$, as follows
\begin{equation}
\begin{array}{l}
\prod\limits_{k = 0}^{N-1} {(1 + \frac{1}{{{\beta _{k,i}}{\mu _i}{\gamma _{k+1,i}}}})} = \\
1+ \frac{1}{{{\mu _i}}}(\sum\limits_{k= 0}^{N-1} {\frac{1}{{{\beta _{k,i}}{\gamma _{k+1,i}}}}} )
+ \frac{{{1 \mathord{\left/
 {\vphantom {1 {2!}}} \right.
 \kern-\nulldelimiterspace} {2!}}}}{{{\mu _i}^2}}(\sum\limits_{k = 0}^{N-1} {\sum\limits_{_{j \ne k}^{j = 0}}^{N-1} {\frac{1}{{{\beta _{k,i}}{\gamma _{k+1,i}}}}} \frac{1}{{{\beta _{j,i}}{\gamma _{j+1,i}}}}} ) +  \cdots\\
+ \frac{{{1 \mathord{\left/
 {\vphantom {1 {N!}}} \right.
 \kern-\nulldelimiterspace} {N!}}}}{{{\mu _i}^N}}(\sum\limits_{k = 0}^{N-1} {\sum\limits_{_{j \ne k}^{j = 0}}^{N-1}{ \cdots \sum\limits_{_{l \ne k,l\ne j,\cdots }^{\hspace{4mm}l = 0}}^{N-1} {\frac{1}{{{\beta _{k,i}}{\gamma _{k+1,i}}}}} \frac{1}{{{\beta _{j,i}}{\gamma _{j+1,i}}}} \cdots \frac{1}{{{\beta _{l,i}}{\gamma _{l+1,i}}}}} } ).
\end{array}
\end{equation}
Then, one can rewrite the rate as
\begin{equation}\label{casl0}
\begin{array}{l}
C=\frac{1}{N}\sum_{i=1}^{N_{F}}\log\left(1+\frac{{\mu_i}^N}{A_{{1},i}{\mu_i}^{N-1}+A_{2,i}{\mu_i}^{N-2}\cdots +A_{N,i} }\right),
\end{array}
\end{equation}
where
\begin{equation}\label{A1}
\begin{array}{l}
{A_{{1},i}} = \frac{1}{{1!}}\sum\limits_{k = 0}^{N-1} {\frac{1}{{{\beta_{k,i}}{\gamma_{k+1,i}}}}} \\
{A_{2,i}} = \frac{1}{{2!}}\sum\limits_{k = 0}^{N-1} {\sum\limits_{_{j \ne k}^{j = 0}}^{N-1} {\frac{1}{{{\beta_{k,i}}{\gamma_{k+1,i}}}}} \frac{1}{{{\beta _{j,i}}{\gamma _{j+1,i}}}}} \\
\hspace{2mm}\vdots \\
{A_{N,i}} = \frac{1}{{N!}}\sum\limits_{k = 0}^{N-1} {\sum\limits_{_{j \ne k}^{j = 0}}^{N-1} { \cdots \hspace{-3mm}\sum\limits_{_{l \ne k,l\ne j,\cdots }^{\hspace{4mm}l = 0}}^{N-1} \hspace{-3mm}{\frac{1}{{{\beta _{k,i}}{\gamma _{k+1,i}}}}} \frac{1}{{{\beta _{j,i}}{\gamma _{j+1,i}}}} \cdots \frac{1}{{{\beta _{l,i}}{\gamma _{l+1,i}}}}} }.
\end{array}
\end{equation}
Under LTIPC, we construct the following Lagrangian function
\begin{equation}\label{opop1}
\begin{array}{l}
\mathcal{L}={\sum_{i=1}^{N_{F}}{\log(1+\frac{{\mu_i}^N}{A_{{1},i}{\mu_i}^{N-1}+A_{2,i}{\mu_i}^{N-2}\cdots +A_{N,i}}}})\\
\hspace{30mm}-\left(\sum_{i=1}^{N_{F}}\lambda_i({\mathbb{E}(\mu_{i})-\frac{1}{N_{F}}})\right)-\boldsymbol{v}^{T}\boldsymbol{\mu},
\end{array}
\end{equation}
where $\boldsymbol{v}^{T}=\left[v_1, \ldots,
v_{N_{F}}\right]$ and $\boldsymbol{\mu}=\left[\mu_1, \cdots  \mu_{N_{F}}\right]$; and $\lambda_i$, and $v_i$ are the Lagrange multipliers corresponding to
the constraints C1 and C2 in \eqref{op2op1}, respectively. Furthermore, by taking the derivative of
$\mathcal{L}$ with respect to ${\mu_i}, i = 1,...,N_{F}$, one
obtains
\begin{equation}\label{eqn1}
(\lambda_i  + \nu_i ){D^2} + ((\lambda_i  + \nu_i ){\mu _i}^N + N{\mu _i}^{N - 1})D - {\mu _i}^N{D^{'}} = 0,
\end{equation}

where
\[\begin{array}{l}
D = {A_{{1},i}}{\mu _i}^{N - 1} + {A_{2,i}}{\mu _i}^{N - 2} \cdots+ {A_{N,i}}\\
\text{and} \quad {D^{'}} = {{\partial D} \mathord{\left/
 {\vphantom {{\partial D} \partial }} \right.
 \kern-\nulldelimiterspace} \partial }{\mu _i}.
\end{array}\]
To find the PA coefficient $\mu_i$, the polynomial
equation in \eqref{eqn1} is to be solved numerically, by considering the power
constraint. The difficulty for solving \eqref{eqn1} increases when the number of hops
increases. For STPC and LTTPC, the procedure is the same as LTIPC, with their corresponding power constraints.

\subsection{The iterative scheme}\label{IPA}
In this subsection, we present an algorithm which iterates between power allocation among relays and subcarriers to maximize the
instantaneous transmission rate of the network. As verified in
Section VI, such an iterative solution can be used as an upper bound
for the performance evaluation of the proposed suboptimal solutions
in Section IV. In this algorithm, optimizations in sub-problem 1 and
sub-problem 2 are repeated alternately, such that the PA
parameters obtained from the previous optimization are the input for
the next one. 
\vspace{2mm}
\begin{tabular}{p{0.2 cm}p{7.5 cm}}\\
\toprule[0.5mm]
& Iterative Algorithm\\
\midrule 1. & Initialize the subcarrier PA coefficients to $\mu_i=1/N_{F}, \forall i= 1,\ldots, N_{F}$. \\
2. &  Given $\mu_i$s, find $\beta_{k,i}$s from the sub-problem 1.\\
3. & Find the instantaneous rate ${C}$ by substituting the PA coefficients $\beta_{k,i}$ and $\mu_i$ in \eqref{casl}.\\
4. & Given $\beta_{k,i}$s, find $\mu_i$s from the sub-problem 2.\\
5. & Find the instantaneous rate ${C}$ in  \eqref{casl} using $\beta_{k,i}$ and $\mu_i$ obtained in steps 2 and 4, respectively.\\
6. & If the difference between rates found in steps 3 and 6 is above a given small value,  repeat steps 2 to 6, if not (or after a predefined number of iterations) go to step 7.\\
7. & Report $\mu_i$ and $\beta_{k,i}$, $\forall i=1,\ldots, N_{F}$ ,$k=1,\ldots, N-1$. \\
\vspace{1mm}\\
\bottomrule[0.5mm]
\end{tabular}
\vspace{5mm}

\section{Power Allocation Schemes in High-SNR Regime}\label{asy}

In this Section, we focus on a high SNR regime in the multi-hop
system and present efficient solutions for the power allocation problem in (6).
The motivation to study such power allocation algorithms is their simplicity with
respect to the iterative solution presented in the previous section.
We start by rewriting the instantaneous rate expression in
\eqref{casl} as
\begin{equation}\label{casl1}
\begin{array}{l}
C=\frac{1}{N}\sum_{i=1}^{N_{F}}\log\left(1+\frac{1}{\frac{A_{{1},i}}{\mu_i}+\frac{A_{2,i}}{{\mu_i}^{2}}+\cdots +\frac{A_{N,i}}{{\mu_i}^{N}} }\right).
\end{array}
\end{equation}
In high SNR regime, the parameters $A_{k,i}$ in (17), for
$k=2,\cdots, N$, are negligible in comparison with $A_{{1},i}$, hence
we neglect higher order terms and rewrite the expression in
\eqref{casl1} as follows
\begin{equation}\label{casl5}
\begin{array}{l}
C_{app}= \frac{1}{N}\sum_{i=1}^{N_{F}}\log\left(1+\frac{\mu_i}{A_{{1},i} }\right).\\
\end{array}
\end{equation}
Substituting $A_{{1},i}$ from \eqref{A1}, the instantaneous rate
is expressed as
\begin{equation}\label{casl4}
\begin{array}{l}
C_{app}=\frac{1}{N}\sum_{i=1}^{N_{F}}\log(1+\frac{\mu_i}{\sum\limits_{k = 0}^{N-1} {\frac{1}{{{\beta _{k,i}}{\gamma _{k+1,i}}}}}}).\\
\end{array}
\end{equation} 
Using \eqref{casl4}, iterative and non-iterative power allocation schemes are developed in the following subsections.
\subsection{Iterative PA in high-SNR regime} \label{asy1}

In high-SNR regime, the steps 2 and 4 in the iterative algorithm can be
implemented in a simpler way, as described below.

\emph{A.1 Sub-problem 1: power allocation among relays}

According to \eqref{casl4}, the
instantaneous rate of the $i$th subcarrier is given by
\begin{equation}\label{casl3}
\begin{array}{l}
{C_i}_{app}=\log(1+\frac{\mu_i}{\sum\limits_{k = 0}^{N-1} {\frac{1}{{{\beta _{k,i}}{\gamma _{k+1,i}}}}}}).\\
\end{array}
\end{equation}
By formulating the power allocation problem for multi-hop narrowband system, the general
optimization problem is written as follows:
\begin{equation}\label{optproblemsch}
\begin{array}{l}
\hspace{10mm}\max_{\{\beta_{k,i}\}}  {C_i}_{app} \\
\mathrm{s.t.}\\
\hspace{5mm}\sum_{k=0}^{N-1}\beta_{k,i}=1\hspace{40.5mm} \hspace{5mm}\text{STPC} \\
\hspace{5mm}\mathbb{E}(\beta_{k,i})=\frac{1}{N},\quad k=0, 1, \ldots, N-1\hspace{15.5mm} \text{LTIPC} \\
\hspace{5mm}\sum_{k=0}^{N-1}\mathbb{E}(\beta_{k,i})=1\hspace{40.5mm} \text{LTTPC} \\
\end{array}
\end{equation}
which is a convex power allocation problem and using the Lagrange method, the
PA coefficients under long-term individual power constraint are derived as
\begin{equation}\label{be1}
\begin{array}{l}
{\beta _{k,i}} = \frac{1}{{N\sqrt {\frac{{\pi {\gamma _{k + 1,i}}}}{{{\Gamma _{k + 1,i}}}}} }} = \frac{{\left| {{\sigma _{k + 1,i}}} \right|}}{{N\sqrt \pi  \left| {{a_{k + 1,i}}} \right|}}.
\end{array}
\end{equation}
And, the power allocation coefficients under LTTPC are calculated as
\begin{equation}\label{be2}
\begin{array}{l}
{\beta _{k,i}} = \frac{1}{{\sqrt \pi  \sum\limits_{j = 1}^N {\sqrt {\frac{{{\gamma _{k + 1,i}}}}{{{\Gamma _{j,i}}}}} } }} = \frac{1}{{\sqrt \pi  \sum\limits_{j = 1}^N {\frac{{\left| {{a_{k + 1,i}}} \right|}}{{\left| {{\sigma _{j,i}}} \right|}}} }}.
\end{array}
\end{equation}
Moreover, the power allocation coefficients under STPC are derived as
follows
\begin{equation}\label{be2}
\begin{array}{l}
{\beta _{k,i}} =\frac{1}{\sum\limits_{j=1}^{N}\sqrt{\frac{\gamma_{k+1,i}}{\gamma_{j,i}}}}=\frac{1}{\sum\limits_{j=1}^{N}\abs{\frac{\alpha_{k+1,i}}{\alpha_{j,i}}}}.
\end{array}
\end{equation}
Note that the PA solutions in \eqref{be1}-\eqref{be2}
were first derived in \cite{poaf}, where the authors investigated high-SNR power allocation
problems for a multi-hop narrowband system model. However, here we use such solutions to provide power allocation
scheme for a wideband multi-hop system with OFDM modulation.

\emph{A.2 Sub-problem 2: power allocation among subcarriers}

In high-SNR regime, the optimization problem in \eqref{op2op1} is
rewritten as
\begin{equation}\label{opt42news}
\begin{array}{l}
\hspace{6mm}\max_{\{{\mu_i}\}} \quad  C_{app} \quad  \mathrm{s.t.}\\
\mathrm{C1.} \\
\hspace{5mm}\sum_{i=1}^{N_{F}}\mu_{i}=1\hspace{15mm} \text{STPC}\\
\hspace{5mm}\mathbb{E}(\mu_{i})=\frac{1}{N_{F}}\hspace{15.5mm} \text{LTIPC}\\
\hspace{5mm}\sum_{i=1}^{N_{F}}\mathbb{E}(\mu_{i})=1\hspace{10mm} \text{LTTPC}\\

\mathrm{C2.} \quad \mu_i \geq 0 \quad  i= 1,\ldots, N_{F}.
\end{array}
\end{equation}
where $C_{app}$ is given in \eqref{casl4}. Since the objective
function in \eqref{opt42news} is concave (see Appendix
\ref{ap2conc}) and the constraints are linear, the optimization
problem (23) has a unique optimal solution. 

Under LTIPC, one can construct the Lagrangian function as
follows
\begin{equation}\label{opop1}
\begin{array}{l}
\mathcal{L}={\sum_{i=1}^{N_{F}}{\log(1+\frac{\mu_i}{\sum_{k=0}^{N-1}\frac{1}{\gamma_{k+1,i}\beta_{k,i}}}}})\\
\hspace{30mm}-\left(\sum_{i=1}^{N_{F}}\lambda_i({\mu_i-\frac{1}{N_{F}}})\right)-\boldsymbol{v}^{T}\boldsymbol{\mu},
\end{array}
\end{equation}
where $\lambda_i$, $i=1, 2, \ldots,N_{F}$, and the vector
$\boldsymbol{v}^{T}=\left[v_1, \ldots, v_{N_{F}}\right]$
are Lagrange multipliers. The optimized $\mu_i$ is
calculated by setting $\frac{\partial\mathcal{ L}}{\partial \mu_i}
=0$, as follows
\begin{equation}\label{most1}
\begin{array}{l}
\mu_i=\left[\frac{1}{\lambda_i}-\sum_{k=0}^{N-1}\frac{1}{\gamma_{k+1,i}\beta_{k,i}}\right]^{+}.
\end{array}
\end{equation}
In (25), the constant $\lambda_i$ is found to satisfy the
constraint C1 in \eqref{opt42news} with equality, i.e.,
\begin{equation}
\begin{array}{l}
\mathbb{E}\bigg(\left[\frac{1}{\lambda_i}-\sum_{k=0}^{N-1}\frac{1}{\gamma_{k+1,i}\beta_{k,i}}\right]^{+}\bigg)=\frac{1}{N_{F}}
\end{array}
\end{equation}
or
\begin{equation}\label{39}
\begin{array}{l}
\int_0^{\frac{{{1}}}{{  {\lambda _i}}}} {{f_Y}(y)dy} =\frac{1}{N_{F}}
\end{array}
\end{equation}
where
\begin{equation}
\begin{array}{l}
Y=\sum_{k=0}^{N-1}\frac{1}{\gamma_{k+1,i}\beta_{k,i}}.
\end{array}
\end{equation}
In \eqref{39}, $f_Y(y)$ is the probability density function (PDF) of $Y$. The PDF of $Y$ is calculated by the convolution of probability density functions for $\frac{1}{\gamma_{k+1,i}\beta_{k,i}}$, $k\in\{1,\cdots,N-1\}$, where $\gamma_{k+1,i}$
follows an exponential distribution, and
$\beta_{k,i}$ is a given constant.

Using the same procedure, the solution for this sub-problem 2, under the LTTPC is derived as
\begin{equation}\label{most2}
\begin{array}{l}
\mu_i=\left[\frac{1}{\lambda}-\sum_{k=0}^{N-1}\frac{1}{\gamma_{k+1,i}\beta_{k,i}}\right]^{+}
\end{array}
\end{equation}
where the constant $\lambda$ is found to satisfy the constraint C1 in \eqref{opt42news} with equality, i.e.,
\begin{equation}
\begin{array}{l}
\sum\limits_{i=1}^{N_{F}}\mathbb{E}\bigg(\left[\frac{1}{\lambda}-\sum_{k=0}^{N-1}\frac{1}{\gamma_{k+1,i}\beta_{k,i}}\right]^{+}\bigg)=1.
\end{array}
\end{equation}
The procedure for STPC is the same as LTTPC.

\subsection{Non-iterative PA scheme in high-SNR regime using channel statistics}
However in previous sections we have presented two PA schemes which utilize CSI for power allocation  among subcarriers and relays, low complexity schemes which can work with channel statistics instead of the CSI are always appreciated. As we saw in Section \ref{asy1}, specially in \eqref{be1}-\eqref{be2}, power allocation among relays is independent from power allocation
among subcarriers. This fact motivates us to present a non-iterative power allocation
algorithm in high-SNR regime. To this end, we insert the derived $\beta_i$:s in \eqref{be1}-\eqref{be2} into \eqref{most1} and \eqref{most2}.
Then, the PA coefficient ${\alpha _{k,i  }}={\mu
_i}{\beta _{k,i}}$ is derived by substituting the instantaneous values with their means. In this case, the power allocation coefficient under long-term total power constraint will be as:
\[{\alpha _{k,i}} = {\left[ {\frac{1}{\lambda } - \sum\limits_{k = 0}^{N-1} {\frac{{\sqrt \pi  }}{{\sqrt {{\gamma _{k + 1,i}}} }}\sum\limits_{j = 1}^N {\frac{1}{{\sqrt {{\Gamma _{j,i}}} }}} } } \right]^ + }{\left[ {\sum\limits_{j = 1}^N {\sqrt {\frac{{\pi {\rm{ }}{\gamma _{k + 1,i}}}}{{{\Gamma _{j,i}}}}} } } \right]^{ - 1}}.\]
For the case of long-term individual power constraint, the solution is given by
\[{\alpha _{k,i}} = \begin{array}{*{20}{l}}
{{{\left[ {\frac{1}{{{\lambda _i}}} - \sum\limits_{k = 0}^{N-1} {\frac{{N\sqrt \pi  }}{{\sqrt {{\gamma _{k + 1,i}}{\Gamma _{k + 1,i}}} }}} } \right]}^ + }}
\end{array}\frac{{\sqrt {{\Gamma _{k + 1,i}}} }}{{N\sqrt {\pi {\rm{ }}{\gamma _{k + 1,i}}} }}.\]
Moreover, by considering a short-term power constraint, we obtain
\[{\alpha _{k,i}} = {\left[ {\frac{1}{\lambda } - \sum\limits_{k = 0}^{N - 1} {\frac{1}{{\sqrt {{\gamma _{k + 1,i}}} }}\sum\limits_{j = 1}^N {\frac{1}{{\sqrt {{\gamma _{j,i}}} }}} } } \right]^ + }{\left[ {\sum\limits_{j = 1}^N {\frac{{{\gamma _{k + 1,i}}}}{{{\gamma _{j,i}}}}} } \right]^{ - 1}},\]
where $\lambda$ and $\lambda_i$ are found using STPC, LTTPC, and
LTIPC power constraints in \eqref{optproblemsch}, respectively. In
the next section we evaluate the performance of proposed power allocation
algorithms.


\section{Performance Evaluation}
\begin{figure}[tbp]
\centering
  \includegraphics[width=7in]{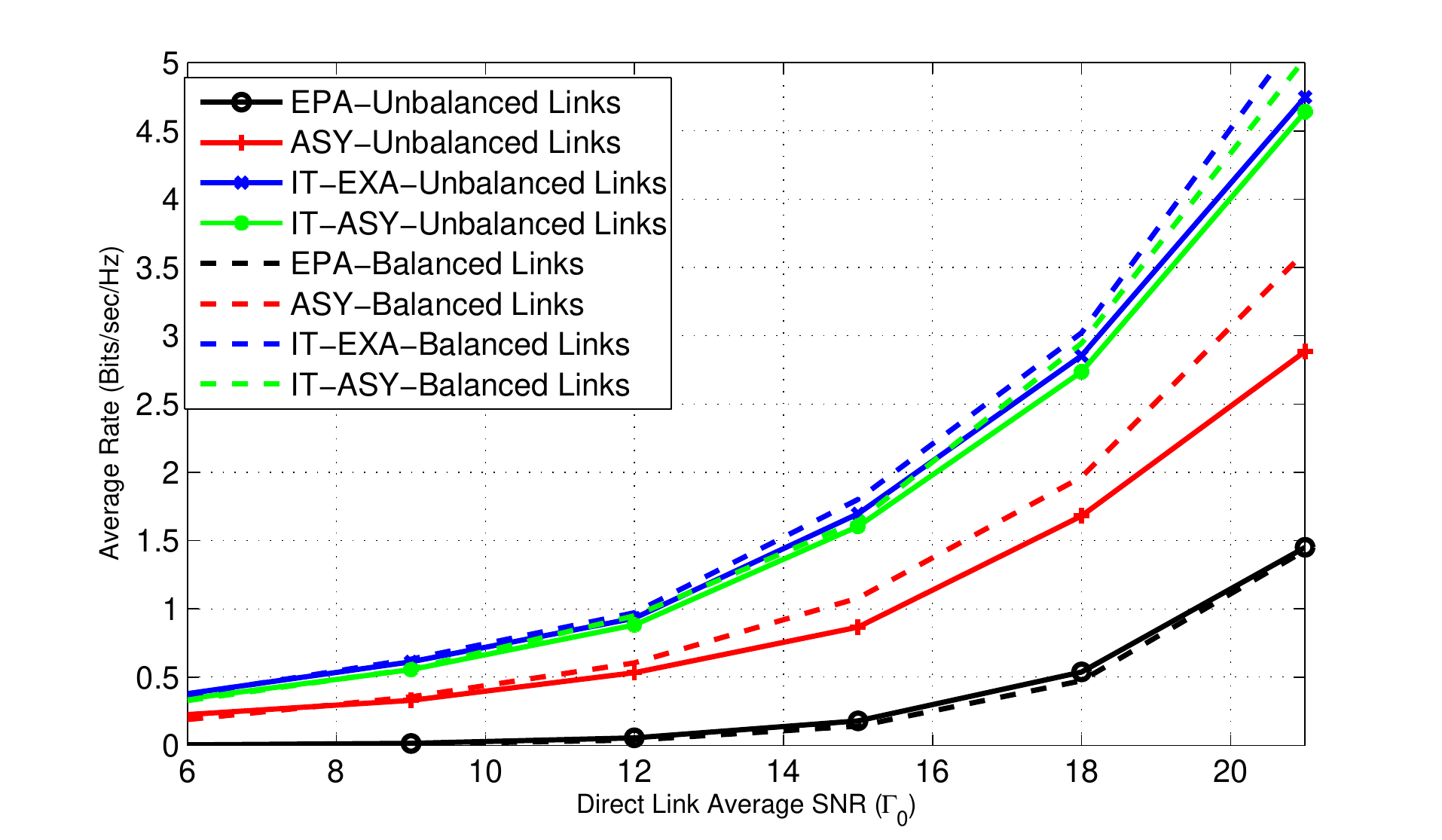}
  \caption{Average rate of 2-hop OFDM ($N_{F}=64$) relaying system  under LTIPC. }\label{fig2}
\end{figure}

This Section presents simulation results for performance evaluation
of the proposed power allocation schemes. In simulations, we assume a multi-hop AF
relaying system over Rayleigh fading channels under short and
long-term individual and total power constraints. In all following
figures, EPA, ASY, IT-EXA, and  IT-ASY refer, respectively, to the
equal PA method, the PA scheme using high-SNR analysis (Section IV),
the iterative power allocation algorithm using exact SNR analysis (Section III),
and the iterative power allocation algorithm using high-SNR analysis (Section
IV). \\
Fig. \ref{fig2} shows the average rate of the 2-hop OFDM relaying
system with $N_{F}=$64 subcarriers versus average SNR of the
direct link $\Gamma_0$, under long-term total power constraints.
Balanced and unbalanced links are considered here. For modeling
unbalanced links in multi-hop system, we adopt the setup of
\cite{poaf}, in which it is assumed that the $k$th terminal, $k =
1,..., N+1,$ is located at the distance $d_k =
\frac{2k}{(N+1)(N+2)}d$ from its previous terminal, where $d$ is the
distance between source and destination. Hence, using the Friis
propagation formula \cite{rap}, the average SNR of $k$th hop is
given by $\Gamma_k=(\frac{(N+1)(N+2)}{2k})^{\delta}\Gamma_0$, where
$\delta$ is the path loss exponent and $\Gamma_0$ is the average SNR
of direct link. We consider $\delta=4$ in this work. For balanced
links, the inter-distance among nodes is considered the same, then the average SNR of $k$th hop is
given by $\Gamma_k=(N+1)^{\delta}\Gamma_0$.

In Fig. \ref{fig2}, one can see that the iterative scheme (IT-EXA) acts as an
upper-bound for other power allocation schemes. The iterative scheme using
high-SNR analysis (IT-ASY) has acceptable performance with a lower
computational complexity because we have closed-form expressions for $\beta_i$:s and $\mu_i$:s in IT-ASY scheme. Also the non-iterative scheme using high-SNR
analysis (ASY) shows a superior performance with respect to equal
power allocation, however, the difference between ASY and IT-EXA
scheme is quite large because we use channel statistics instead of the CSI in ASY scheme.

\begin{figure}[tbp]
\centering
\includegraphics[width=7in]{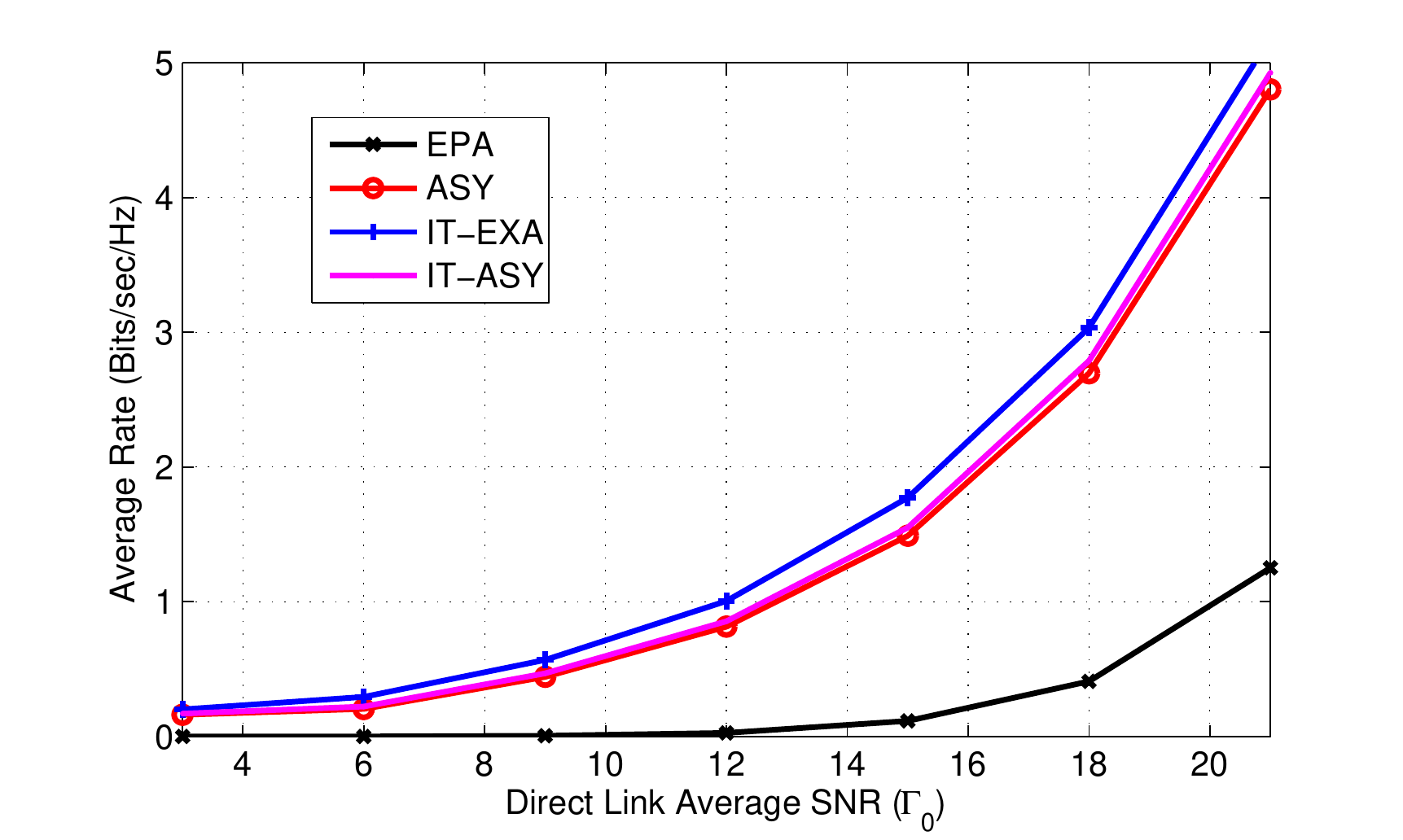}
  \caption{Average rate of 3-hop OFDM ($N_{F}=64$) relaying system  under STPC. }\label{fig3}
\end{figure}

\begin{figure}[tbp]
\centering
\includegraphics[width=7in]{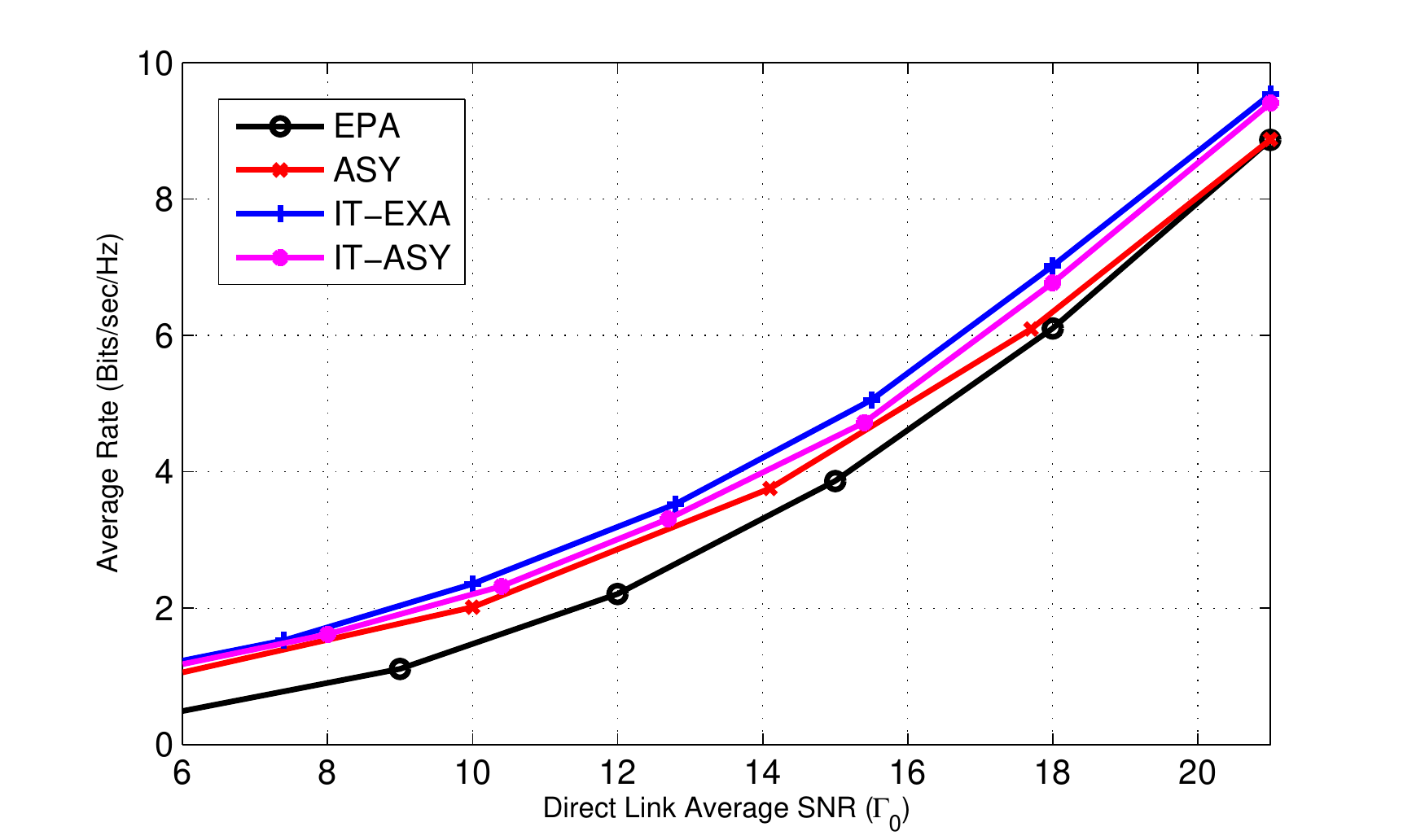}
  \caption{Average rate of 2-hop OFDM ($N_{F}=8$) relaying system  under LTTPC. }\label{fig35}
\end{figure}

Fig. \ref{fig3} shows the average rate of the 3-hop OFDM relaying
system with 64 subcarriers versus average SNR of the direct link, $\Gamma_0$, under short-term power constraints. One can see that the
performance of iterative algorithm based on exact SNR analysis is
superior than other schemes. Moreover, the performance of the
non-iterative power allocation scheme is close to that of the low complexity
iterative algorithm, and both of these schemes provide significant
gain over the equal power allocation. The same results are depicted
in Fig. \ref{fig35} for 2-hop system under long-term total power
constraint.

\begin{figure}[tbp]
\centering
\includegraphics[width=7in]{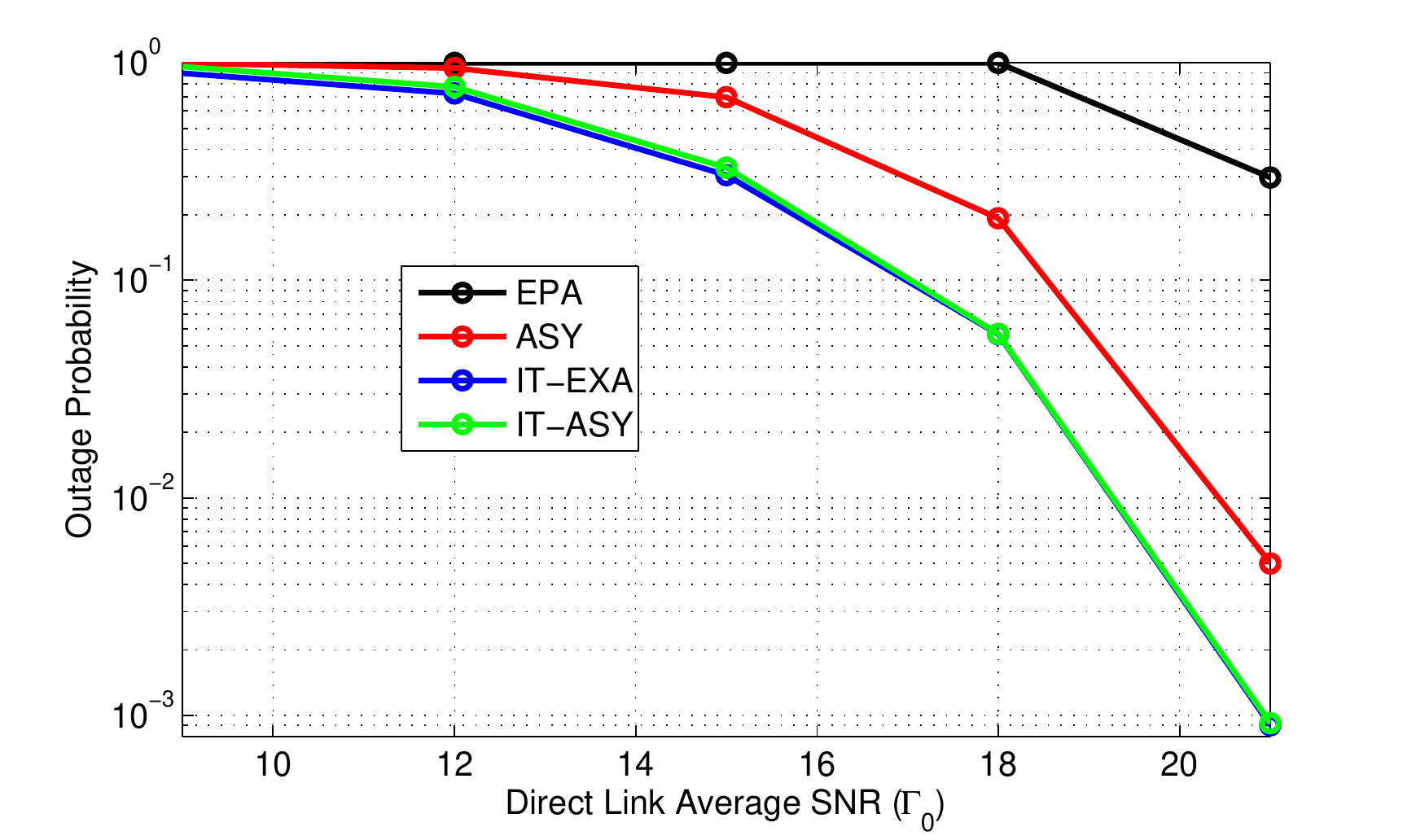}
  \caption{Outage probability of 2-hop OFDM ($N_{F}=64$) relaying system  under LTIPC.  }\label{fig4}
\end{figure}

\begin{figure}[tbp]
\centering
\includegraphics[width=7in]{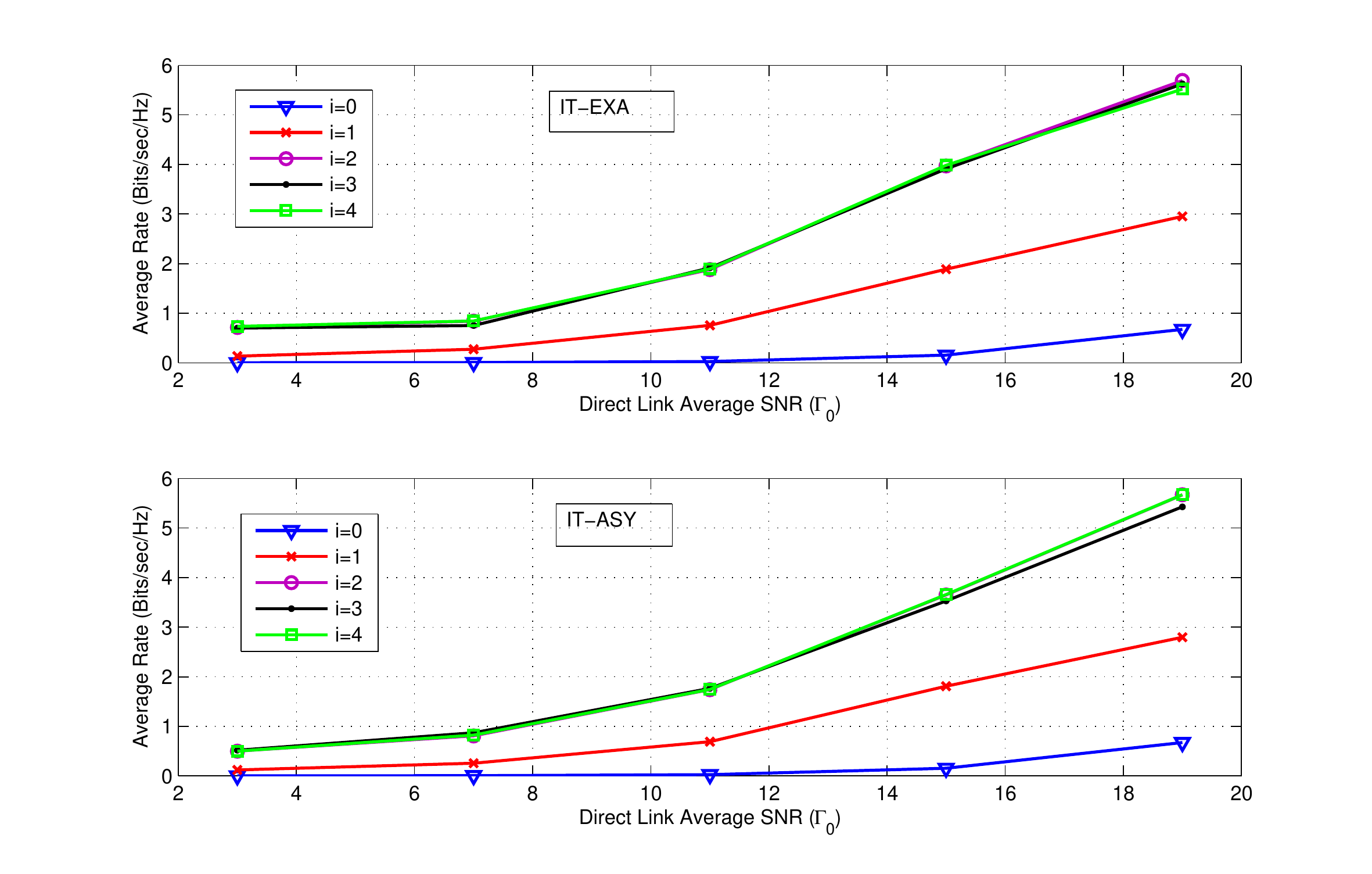}
  \caption{Convergence of iterative algorithm (exact and asymptotic) for PA in 2-hop OFDM ($N_{F}=64$) relaying system. $i$ stands for the number of iterations.  }\label{fig44}
\end{figure}

In Fig. \ref{fig4}, the outage probability of 2-hop OFDM system
($N_{F}=64$) is depicted versus the average SNR of direct link,
$\Gamma_0$, for several power allocation schemes under long-term
individual power constraints. The outage probability is defined as
the probability that the instantaneous rate falls below 1
bit/sec/Hz. In Fig. \ref{fig44}, the average rate of 2-hop OFDM
system when iterative PA algorithms in both exact and asymptotic
forms are applied (under LTTPC) is depicted. In this figure, the
index $i$ stands for the number of iterations considered, while
$i=0$ refers to the uniform power allocation scenario. As an
interesting observation, from this figure we can see that after a
few iterations the average rate converges to its maximum value. We
also note that from a complexity perspective, the computational
complexity of proposed iterative algorithms is directly related to
the complexity of sub-problems in each iteration. As an example, for
the 2-hop OFDM system considered in Fig. \ref{fig44}, the complexity
of IT-ASY and IT-EXA algorithms may be easily related to the
complexity of solving PA sub-problems among subchannels and relays
(the corresponding waterfilling solutions for these sub-problems are
presented in Section III and IV). In particular, the complexity of
waterfilling solutions has been already investigated in
\cite{compx}.

From Fig. \ref{fig2}-\ref{fig4} we make the following observations:
(i) The IT-EXA scheme provides the best performance among other
methods at the cost of a higher computational complexity; (ii) The
IT-ASY scheme provides a data rate performance which is very close
to that of the IT-EXA scheme, while it enjoys a considerable lower
complexity; (iii) The ASY scheme provides an acceptable level of
data rate performance due to its considerable lower complexity and
easier implementation as it needs channel statistics instead of CSI; (iv) The
performance of the IT-ASY scheme in high SNR regime converges to
that of the IT-EXA scheme, as it is verified in Fig. \ref{fig35} and
Fig.\ref{fig4}; (iv) For a multi-hop OFDM scenario, the proposed
power allocation schemes greatly outperform the scheme with uniform
power allocation.


\section{Conclusion}
In this paper we considered the problem of power allocation in
narrowband and broadband (OFDM) multi-hop relaying systems employing
non-regenerative relays with different power constraints. We
proposed exact and approximate design approaches depend on the
wireless application demand and network structure. In particular,
aiming at maximizing the instantaneous multi-hop transmission rate,
several power allocation algorithms have been developed in a unified framework
including: (i) an iterative power allocation method which provides an upper-bound
performance; (ii) a relatively low-complexity iterative power allocation method;
and (iii) a non-iterative power allocation scheme with acceptable performance at
high SNR regime. Moreover, we provided performance comparison with
respect to an equal power PA solution and quantify the rate
performance loss incurred at the price of low complexity and low
feedback overhead.

\appendices

\section{}\label{ap1conv}
Here, we prove that the objective function in \eqref{optschaver1} is convex. Let
denote this function by f($\beta_k$) as follows
\begin{equation}
\begin{array}{l}
\text{f}(\beta_k)=\sum_{k=1}^{N-1}{\ln(1+\frac{1}{\beta_k\gamma_{k+1}})}.
\end{array}
\end{equation}
We can easily obtain
\begin{equation}\label{43}
\begin{array}{l}
\frac{{\partial}^2 \text{f}(\beta_k)}{{\partial
\beta_k}^2}=\frac{1+2\beta_k
\gamma_{k+1}}{{(\beta_k(\beta_k\gamma_{k+1}+1))}^2}.
\end{array}
\end{equation}
Since the coefficient $\beta_k$ is between 0 and 1 (see Section
II), and $\gamma_{k+1}$ takes positive values, the second derivative in \eqref{43} is positive for
any channel realization, and f($\beta_k$) is convex.

\section{}\label{ap2conc}
Here, we show that the objective function in \eqref{opt42news} is
concave. Let f($\mu_i$) denote the objective function, that is
\begin{equation}
\begin{array}{l}
\text{f}(\mu_i)=\sum_{k=1}^{N_{F}}{\log\left(1+\frac{\mu_i
P_{T}}{N\sqrt{\pi}\sum_{k=1}^{N}{\frac{{N_0}_{k,i}}{\abs{a_{k,i}}\sigma_{k,i}}}}\right)}.
\end{array}
\end{equation}
We rewrite this function as follows
\begin{equation}
\begin{array}{l}
\text{f}(\mu_i)=\sum_{k=1}^{N_{F}}{\log\left(1+\mu_i g_i\right)}
\end{array}
\end{equation}
where
\begin{equation}\label{g}
\begin{array}{l}
\text{g}_i=\frac{P_{T}}{N\sqrt{\pi}\sum_{k=1}^{N}{\frac{{N_0}_{k,i}}{\abs{a_{k,i}}\sigma_{k,i}}}}.
\end{array}
\end{equation}
Taking the second derivative with respect to $\beta_k$, one
obtains
\begin{equation}\label{47}
\begin{array}{l}
\frac{{\partial}^2 \text{f}(\mu_i)}{{\partial
\mu_i}^2}=-\frac{\log(e){g_i}^2}{{1+\mu_i g_i}^2}
\end{array}
\end{equation}
As stated in Section II, $\mu_i$ is between 0 and 1 for any
channel realization. Hence, the second derivative in \eqref{47} is
negative for any channel realization, and the function f($\mu_i$) is
concave.

\section*{Acknowledgment}
A. Azari would like to thank H. Khodakarami, R. Hemmati, A. Behnad, and R. Parseh
for their helpful discussions on this work.

\ifCLASSOPTIONcaptionsoff
  \newpage
\fi

\bibliographystyle{IEEEtran}
\bibliography{IEEEabrv,paper_re}

\end{document}